\documentclass[aps,floatfix,prd,showpacs,twocolumn]{revtex4}

\usepackage{graphicx}
\usepackage{dcolumn}
\usepackage{bm}

\begin{document}

\def\be{\begin{equation}}
\def\ee{\end{equation}}
\def\bd{\begin{displaymath}}
\def\ed{\end{displaymath}}
\def\ba{\begin{eqnarray}}
\def\ea{\end{eqnarray}}
\def\lr{\leftrightarrow}
\def\cc{$c\bar c$ }
\def\ccbar{$c\bar c$ }
\def\3P0{$^3{\rm P}_0$ }
\def\x{\bf x}
\def\h{\rm h}
\def\A{\rm A}
\def\B{\rm B}
\def\C{\rm C}
\def\D{\rm D}
\def\H{\rm H}
\def\I{\rm I}
\def\J{\rm J}
\def\E{\rm E}
\def\L{\rm L}
\def\K{\rm K}
\def\M{\rm M}
\def\N{\rm N}
\def\P{\rm P}
\def\S{\rm S}
\def\Db{\bar{\rm D}}
\def\Pb{\bar{\rm P}}
\def\Sb{\bar{\rm S}}

\newcommand{\sixj}[6]{ \left\{\begin{array}{ccc}  {#1} & {#2} & {#3} \\    {#4}
& {#5} & {#6} \end{array}   \right\} }
\newcommand{\ninej}[9]{ \left\{\begin{array}{ccc}  {#1} & {#2} & {#3} \\    {#4}
& {#5} & {#6}\\    {#7} & {#8} & {#9}  \end{array}   \right\} }

\title{Hadron Loops: General Theorems\\ and Application to Charmonium} 
\author{
T. Barnes$^a$\footnote{Email: tbarnes@utk.edu}
and
E.S. Swanson$^b$\footnote{Email: swansone@pitt.edu}
}
\affiliation{
$^a$
Department of Physics and Astronomy, 
University of Tennessee,
Knoxville, TN 37996,
USA,\\
Physics Division, Oak Ridge National Laboratory,
Oak Ridge, TN 37831, USA\\
$^b$Department of Physics and Astronomy, 
University of Pittsburgh, 
Pittsburgh, PA 15260, 
USA.}

\date{\today}

\begin{abstract}
In this paper we develop a formalism for incorporating 
hadron loops in the quark model. We~derive expressions for mass shifts, continuum 
components and mixing amplitudes of ``quenched" quark model states due to hadron loops, 
as perturbation series in the valence-continuum coupling Hamiltonian. We prove three general theorems 
regarding the effects of hadron loops, which show that given certain
constraints on the external ``bare" quark model states, the valence-continuum coupling, 
and the hadrons summed in the loops, the following results hold:
(1) The loop mass shifts are identical for all states within a given N,L multiplet. 
(2) These states have the same total open-flavor decay widths. 
(3) Loop-induced valence configuration 
mixing vanishes provided that ${\L}_i \neq \L_f$ or $\S_i \neq \S_f$. 
The charmonium system is used as a numerical case study, with 
the $^3\P_0$ decay model providing the valence-continuum coupling. 
We evaluate the mass shifts and continuum mixing numerically 
for all 1S, 1P and 2S charmonium valence states due to
loops of D, D$^*$, D$_s$ and D$_s^*$ meson pairs. 
We find that the mass shifts are quite large, but are numerically
similar for all the low-lying charmonium states, as suggested by the first theorem.
Thus, loop mass shifts may have been ``hidden" in the valence quark
model by a change of parameters. 
The two-meson continuum components of the physical charmonium states are also found to 
be large, creating challenges for the interpretation of the constituent quark model.

\end{abstract}

\maketitle

\section{Introduction}

The discovery of the narrow charm-strange mesons 
D$_{s0}^*$(2317)$^+$  
\cite{Aubert:2003fg}
and
D$_{s1}$(2460)$^+$  
\cite{Besson:2003cp}
has given special impetus to the calculation of hadron loop effects,
since the loops are often cited as a possible reason for the 
surprisingly low masses of these mesons.
This possibility is supported by the prediction that the corresponding $c\bar s$ 
quark model states have especially strong couplings to the open-flavor decay channels
DK and D$^*$K \cite{Godfrey:wj,Close:2005se}.
(For discussions of the importance of hadronic loops effects in this and
other contexts, including other heavy-quark mesons, see for example 
Refs.\cite{Swanson:2005rc, Hwang:2004cd, Eichten:2004uh, Eichten:1979ms, Eichten:1978tg, 
Eichten:ag, Eichten:1975bk, Rupp:2006sb, van Beveren:2006st, 
van Beveren:2006ih, van Beveren:2005ha, van Beveren:2004ve,
van Beveren:2004bz, van Beveren:2003af, van Beveren:2003jv, 
van Beveren:2003kd, van Beveren:qb, van Beveren:bd, 
Hanhart:2007yq, Kalashnikova:2007qz, Kalashnikova:2005ui, 
Amsler:2004ps, Ono:1983rd, Heikkila:1983wd, Pennington:2007xr, 
Geiger:va, Geiger:qe, Geiger:ab, Geiger:yc, Morel:2002vk, Zenczykowski:1985uh, Torn,
Tornqvist:1984fy}.)

The subject of valence-continuum couplings is also relevant to the  
X$(3872)$ seen in $J/\psi \pi^+ \pi^-$ \cite{Choi:2003ue,Acosta:2003zx}, 
which may be dominantly a weakly-bound $1^{++}$ DD$^*$ molecular state; the size of the 
$c\bar c$ valence component present in this system has important implications 
for the properties of this state. (See Ref.\cite{Swanson:2006st} for a review 
of these and other recent developments in heavy-flavor hadrons.)

Since the open-flavor decay couplings of hadrons 
to two-body final states 
$\A\to \B\C$ are large, one might anticipate that 
second-order decay loops, in which a hadron virtually decays to
a two-body intermediate state and then reforms the original hadron
($\A\to \B\C\to \A$), are also important effects. 
These second-order virtual processes give rise to mass shifts 
of the bare hadron states, and contribute continuum components 
to the physical hadron state vectors. A careful estimate of these 
mass shifts is of great interest, since they are usually not 
included in quark potential models 
and are only partially present in 
quenched lattice QCD, 
and may constitute important ``systematic" errors in the results.

In our initial study we develop a formalism for treating 
these loops effects, using results from our earlier studies of
open-flavor decay amplitudes. In particular we give results for the 
loop-induced mass shifts and continuum amplitudes of hadrons, as well as the
off-diagonal ``spectroscopic mixing amplitudes" induced by hadron loops between different 
external discrete hadron basis states. 

As a numerical application we 
consider the charmonium system, and evaluate these mass shifts
and continuum components for the lighter (narrow) charmonium states 
that lie below the open charm threshold.
Charmonium is especially attractive as a test system for studying
loop effects because the low-lying spectrum is clear experimentally, 
with complete 1S, 1P and 2S multiplets, and all eight states in 
these multiplets are below the open-flavor decay threshold. 
(This implies that all mass shifts are negative, with no cancellations.) 
In addition the charmonium system is only moderately relativistic, and 
the spectrum is quite well described by quenched potential
models and lattice gauge theory. Thus loop effects may be
evaluated as (possibly) perturbative corrections to well understood 
$c\bar c$ potential model states, and the results may be contrasted
with an unambiguous experimental spectrum.

\section{Formalism}

\subsection{The Loop Model}
                                                                                                                
To incorporate hadron loop effects in the quark model,
we model a physical hadron as a bare valence state
$|{\A}\rangle$ augmented by
two-hadron continuum components, 
\be
|\Psi\rangle =
|{\A} \rangle
+
\sum_{\B\C}
\psi_{\B\C} |{\B\C} \rangle \ .
\ee
We assume that the Hamiltonian for this combined system  
consists of a valence Hamiltonian ${\H}_0$ (the 
quark model Hamiltonian) and an interaction 
${\H}_{\I}$ which couples the valence and continuum sectors,
\be
{\H} = {\H}_0 + {\H}_{\I}\ .
\ee

We will evaluate the continuum components of the hadron state and
their physical effects as a perturbation series in the 
valence-continuum coupling ${\H}_{\I}$. 
Our starting point for this perturbations series is the set
of single valence hadron ${\H}_0$ eigenstates; a specific valence 
state is written as 
$|{\A}({\vec p}_{\A})\rangle$, and is assigned an ${\H}_0$ eigenvalue of
${\E}_{\A} = ({\M}_{\A}^2 + {\vec p}_{\A}^{\; 2})^{1/2}$.
Since we normally work in the rest frame of the valence hadron,
${\vec p}_{\A} = 0$, this energy eigenvalue is just 
the rest mass ${\M}_{\A}$ of the bare valence quark model hadron. 

The free two-hadron valence states which form our 
zeroth-order noninteracting continua are written as 
$|{{\B}({\vec p}_{\B}){\C}({\vec p}_{\C})}\rangle$. 
The valence Hamiltonian ${\H}_0$ is understood to operate 
only between the constituents of B and C separately; 
BC interactions, which are not treated here, would be 
incorporated in a separate two-hadron interaction Hamiltonian.  
This BC continuum state has ${\H}_0$ eigenvalue
${\E}_{\B\C} = {\E}_{\B} + {\E}_{\C}$ 
where
${\E}_{\B} = ({\M}_{\B}^2 + {\vec p}_{\B}^{\; 2})^{1/2}$
and
${\E}_{\C} = ({\M}_{\C}^2 + {\vec p}_{\C}^{\; 2})^{1/2}$.
In the A rest frame we have 
${\vec p}_{\B} = - {\vec p}_{\C} \equiv {\vec p}$, and
with $p \equiv |{\vec p}\, |$ the energies are
${\E}_{\B} = ({\M}_{\B}^2 + p^2)^{1/2}$
and
${\E}_{\C} = ({\M}_{\C}^2 + p^2)^{1/2}$.

The matrix elements of the valence-continuum coupling 
Hamiltonian are of the form
\be
\langle {\B\C} | {\H}_{\I} |{\A}\rangle = 
h_{fi} \, \delta({\vec p}_{\A} - {\vec p}_{\B} - {\vec p}_{\C}) \ . 
\label{hfi_defn}
\ee
With explicit momentum labels these rest-frame 
one- and two-hadron valence states are written as
$|{\A}({\vec 0})\rangle$ and
$|{\B}({\vec p}\, ){\C}(-{\vec p}\,\, )\rangle$, 
and the coupling matrix element
is a function of a single momentum vector, $h_{fi}({\vec p}\, )$.

\subsection{Mass Shifts}

The mass shift of a valence hadron A due to its coupling to
the BC continuum may be expressed in terms of 
the coupling matrix element
$h_{fi}({\vec p}\, )$ of Eq.(\ref{hfi_defn}) 
using second-order perturbation theory (for a general discussion see
\cite{Fano}).
The usual discrete sum 
$\sum_n$ over intermediate states generalizes to 
a momentum-space integral over continuum states 
$|{\B}({\vec p}\, ){\C}(-{\vec p}\, )\rangle$; the
result for a single BC channel is  
\bd
\Delta {\M}_{\A}^{({\B\C})} \! = 
-\! \sum_n 
 \frac{|\langle \psi_n|{\H}_{\I}|\psi_{{\h}_0}\rangle|^2}{({\E}_n - {\E}_0)}
=
-\!\! \int d^3p\, \frac{| h_{fi}|^2}
{({\E}_{\B\C} - {\M}_{\A})}
\ed
\bd
= - {\cal P} \int_{{\M}_{\B} + {\M}_{\C}}^{\infty}   
\frac{d{\E}_{\B\C}}{({\E}_{\B\C} - {\M}_{\A})} 
\frac{p {\E}_{\B} {\E}_{\C}}{\E_{\B\C}}
\int d\Omega_p | h_{fi}|^2 
\ed  
\be
-i\,\pi\,
\bigg\{ 
\frac{p {\E}_{\B} {\E}_{\C}}{\M_{\A}}
\int d\Omega_p | h_{fi}|^2
\bigg\}
\bigg|_{{\E}_{\B\C} = {\M}_{\A}} 
\label{Mshift}
\ee  
where 
${\cal P}$ is the principal part integral. 
There is an implicit sum 
over any intermediate-state polarization labels 
in the squared Hamiltonian matrix element $|h_{fi}|^2 $. 

As a check of our central result Eq.(\ref{Mshift}),
note that the imaginary part of the mass shift
\be
{\rm Im}(\Delta {\M}_{\A}^{({\B\C})}) 
=
-\pi\,
\bigg\{ 
\frac{p {\E}_{\B} {\E}_{\C}}{{\M}_{\A}}
\int d\Omega_p |h_{fi}|^2 
\bigg\}
\bigg|_{{\E}_{\B\C} = {\M}_{\A}} 
\label{ImE}
\ee  
should be related to the total decay rate by
\be
\Gamma({\A}\to{\B\C}) = -2\, {\rm Im}(\Delta {\M}_{\A}^{({\B\C})}).
\label{ImM_rate}
\ee
The standard $\A \to \B + \C$ decay rate formula 
given in Eq.(5) of Ref.\cite{Ackleh:1996yt} is indeed consistent with this relation. 

If the initial hadron mass 
is below BC threshold (${\M}_{\A} < {\M}_{\B} + {\M}_{\C}$) we do
not encounter a singular energy denominator, and
this mass shift is a real, negative definite  
integral over $p$,
\be
\Delta {\M}_{\A}^{({\B\C})} 
=
 -\int_0^{\infty}  \!\!\! 
\frac{p^2 dp}{({\E}_{\B\C} - {\M}_{\A})}
\int d\Omega_p | h_{fi}|^2\ .
\label{Mshift_pint}
\ee
If one considers mixing between the valence state A and several 
continuum BC channels, the total mass shift at this (leading) order 
in the valence-continuum coupling is the sum of the individual 
mass shifts due to each channel.

Nonperturbative estimates of the mass shift can be made in the  
absence of final state interactions by summing iterated
bubble diagrams. The result is a full propagator of the form

\be
-i G(s) = \frac{1}{(s - \M^2 - \Sigma(s))}
\label{eq:bubble_sum}
\ee
where $\Sigma$ is the one particle irreducible self-energy of the  
meson in question.
The propagator pole 
yields the meson mass shift and width. Contact to our perturbative, 
nonrelativistic results can be made by identifying $\sqrt{s}\, \Gamma(s) =
-{\rm Im}(\Sigma(s))$ and $2 \sqrt{s}\, \delta \M(s) = {\rm Re}(\Sigma(s))$, and 
assuming that the width and mass shift are small relative to the unperturbed 
meson mass.  

We have computed numerical pole positions with iterated loops 
for the charmonium examples of the next section using this formalism, 
and find rather small differences in mass shifts relative to the single 
loop approximation (typically $\Delta \M (c\bar c)$ changes by less than 10\%).

We have also examined the effect of mixing-induced coupling between states. Thus the denominator
of the propagator becomes a matrix, $(s-m_i^2)\delta_{ij} - \Sigma_{ij}(s)$. Solving this equation for
the case of the $J/\psi$ coupling to $\psi'$ and $\psi''$ through $\D\D$, $\D_s\D_s$, $\D\D^*$, 
$\D_s\D_s^*$, 
$\D^*\D^*$, and $\D_s^*\D_s^*$ continua again yields corrections 
to the one-loop diagonal results 
that are typically {\it ca.} 10\%. 
We therefore simply present perturbative (one-loop), single 
channel mass shifts 
in the discussion of charmonium.

\subsection{Continuum Components}

Although mass shifts due to loops may be ``hidden" in fitted parameters in quenched
approaches, such as $m_q$ or $V_0$ in potential models and $m_q$ or $a(\beta)$ in
quenched LGT, it should nonetheless be possible to identify other, more characteristic
effects of the two-meson continuum components. We will require the explicit continuum component 
wavefunctions to evaluate their effects on observables. Here we give general results for 
these wavefunctions; an example will be considered in the discussion of charmonium.

The valence-continuum coupling ${\H}_{\I}$ induces a continuum
component in an initially pure valence state $|{\A}\rangle$. 
At leading order in $h_{fi}$ this continuum component 
is given by
\be
\sum_{\B\C} \psi_{\B\C} |{\B\C}\rangle = 
- ({\H}_0 - {\M}_{\A})^{-1}\; {\H}_{\I}\; |{\A}\rangle \ .
\label{BCcontinuum_state}
\ee
The momentum-space wavefunction of the continuum component 
in a specific channel BC is
\be
\phi_{{\B}(\vec p){\C}(-\vec p)} \equiv \phi_{\B\C}(\vec p \,) =
- \frac{h_{fi}(\vec p\,)}{({\E}_{\B\C}(p) - {\M}_{\A})}\ .
\label{BCcontinuum_wfn}
\ee 
Using the conventions of Ref.\cite{Barnes:1991em}, the corresponding real-space 
wavefunction in the relative separation $\vec r = \vec r_B - \vec r_C$ is
\be
\psi_{\B\C}(\vec r\, )
= \int\! d^{\, 3}p\; \phi_{\B\C}(\vec p \,)\, \frac{{\rm e}^{i \vec p \cdot \vec r}}{(2\pi)^{3/2}}
\ee

For nonzero spin this spatial wavefunction is implicitly summed over the meson 
orbital and spin magnetic quantum numbers,
to give overall states with the $\J,\J_z$ of meson~A. 

The norm of this continuum component gives the 
probability that the physical energy eigenstate
is in the two-meson channel BC. This is
\be
{\P}^{({\B\C})}_{\A} 
=
\sum_n  \frac{|\langle \psi_n |H_I|\psi_0\rangle|^2}{({\E}_n - {\E}_0)^2}
=
\int\! d^{\, 3}p\; \frac{|h_{fi}|^2}
{({\E}_{\B\C} - {\M}_{\A})^2} \ .
\label{BC_prob_0}
\ee
\be
\hskip 2cm
= \int_0^{\infty}  \!\!\!
\frac{p^2 dp}{(E_{BC}-M_A)^2}
\int d\Omega_p | h_{fi}|^2\ .
\label{BC_prob}
\ee

\subsection{Spectroscopic Mixing}

Mixing of discrete ``valence" quark model basis states through hadron loops 
is an interesting effect which may have easily observable consequences.
The amplitude $a_{fi}$ to find a discrete basis state $|f\rangle$ 
in the initially pure valence state $|i\rangle$ as a result 
of continuum mixing is given by second-order perturbation theory,
\be
a_{fi} =
\frac{1}{({\M}_f - {\M}_i)} \sum_{{\B\C}} \int d^{\, 3}p\;
\frac{h_{f,{\B\C}}({\vec p}\, )\,  h_{{\B\C},i}({\vec p}\, )} 
{({\E}_{\B\C}(p) - {\M}_i)} \ .
\label{a_fi}
\ee
For an initial valence state within the continuum this is replaced by a
principal part integral and the amplitude
$a_{fi}$ has an imaginary part, analogous to Eq.\ref{Mshift}.

Note that this loop-induced mixing amplitude is somewhat counterintuitive, 
in that it is nonsymmetric in general;
\be
|\, a_{fi}| \ne |\, a_{if}| \ .
\ee
This disagrees with the simple picture of an orthogonal rotation 
between two basis states often used to describe mixing
in the quark model. (Examples include mixing between spin-singlet and 
spin-triplet axial vector K$_1$ and D$_1$ mesons, and between the  
$|2^3{\S}_1\rangle$
and 
$|{}^3{\D}_1\rangle$ charmonium basis states in the 
$\{ \psi'(3686), \psi(3772)\}$ system.) Since this is actually an
infinite-dimensional Hilbert space rather than a two-dimensional one, 
it is of course not necessary that $|\, a_{fi}| = |\, a_{if}|$. 
Instead the valence state $|i\rangle$ that is closest to the 
continuum, and hence minimizes the valence-continuum energy denominator
$({\E}_{\B\C} - {\M}_i)$ in Eq.(\ref{a_fi}), 
will tend to experience the largest mixing.
This will be illustrated in the next section.

\subsection{Three Loop Theorems}

In the Appendix we show that sums over sets of mesons within the loop under certain 
conditions gives very simple relations between the mass shifts, strong widths, 
and configuration mixing amplitudes due to hadron loops. 
Although these relations are not exactly satisfied in nature, they are
sufficiently accurate to be relevant to realistic problems such as 
the charmonium examples we consider here.

Provided that our conditions are satisfied, one may show that for
the states $\{ {\rm A} \}$ in a given N$_i$,L$_i$ multiplet:
\begin{enumerate}

\item 
The mass shifts for all states $\{ {\rm A} \}$ are equal. 

\item 
Their strong (open-flavor) total widths are equal.

\item 
The configuration mixing amplitude $a_{fi}$ between any two valence
basis states $i$ and $f$ vanishes if ${\L}_i \neq \L_f$ or $\S_i \neq \S_f$.
 
\end{enumerate}
These conclusions hold to all orders if there are no final state interactions in the
continuum channels.

To prove these loop results, we consider a sum over a finite set of intermediate 
(loop) mesons that runs over all mesons in a given N,L multiplet, taking on all 
allowed values of spin S and total angular momentum J. 
Examples of such loop sets include 
``$\S\Sb$"; 
$\{$ $(\D\Db)$, $(\D\Db^*)$, $(\D^*\Db)$, $(\Db^*\Db^*)$ $\}$
and
``$\S\Pb$"; 
$\{$ 
$(\D\Db_0^*)$, $(\D^*\Db_0^*)$, 
$(\D\Db_1)$,   $(\D^*\Db_1)$, 
$(\D{\Db}_1{}^{'})$, $(\D^*{\Db}_1{}^{'})$, 
$(\D\Db_2^*)$, $(\D^*\Db_2^*)$ 
$\}$, where we have explicitly indicated antiparticles.
The proof assumes that all members of the set 
of intermediate (loop) mesons which are summed over have the same mass, 
and for the first two ``diagonal" conclusions (equality of mass shifts 
and total widths) we also require that the external mesons 
have a common bare mass and radial wavefunction. 
In addition there are general conditions that the 
valence-continuum coupling must satisfy, which are discussed 
in the Appendix. These constraints are satisfied by the $^3$P$_0$ model 
that is used for illustration in this paper.

The first conclusion suggests how these intrinsically large loop mass shifts 
can be hidden in the parameters of quenched models; since the largest effect of 
loops on the spectrum of states in the multiplets we consider is an overall 
downward mass shift in the multiplet c.o.g., this can be approximately parameterized
through a change in the quark mass or through a constant in the potential. 

As as illustration of this theorem in a specific case, 
Table~\ref{Table_rel_mass_shifts} shows the relative mass shifts of all the
1P charmonium levels due to their $^3$P$_0$ couplings to 
DD, DD$^*$ and D$^*$D$^*$ 
meson loops, assuming that all the bare 1P $c\bar c$ masses are identical, 
the D and D$^*$ meson masses are identical, and each flavor system has a 
common radial wavefunction.
Although the individual continuum channels DD, DD$^*$ and D$^*$D$^*$ 
make different contributions to the mass
shift of each meson, the {\it summed} mass shift from all three 
channels is identical for each of the four 1P mesons. 
Thus, if the mesons are initially degenerate, they remain degenerate after 
these loop effects are included. 
This has been noted previously by Tornqvist\cite{Torn}; the results presented here can
be considered an elaboration of this original observation. Furthermore, 
Tornqvist and \.{Z}enczykowski have studied the
analogous effect in baryons in Refs. \cite{Tornqvist:1984fy, Zenczykowski:1985uh}.

One may also see evidence for the no-loop-mixing result for states with different
L (conclusion 3 above) in our charmonium example. In Table~\ref{Table_afi} 
we show the individual 
DD, DD$^*$ and D$^*$D$^*$ one-loop contributions to the mixing between 
$|{}^3\S_1\rangle$
$|2{}^3\S_1\rangle$
and
$|{}^3\D_1\rangle$
charmonium valence basis states. Note that in some of the disfavored cases, such as 
$|{}^3\S_1\rangle \to |{}^3\D_1\rangle$, there is an almost complete cancellation 
of the final $|{}^3\D_1\rangle$ amplitude, due to destructive interference between 
the DD, DD$^*$ and D$^*$D$^*$ loops. This destructive interference between 
loops is still evident but less complete for mixing between the higher-lying states 
$|2{}^3\S_1\rangle$
and
$|{}^3\D_1\rangle$,
because they are quite close to DD threshold; this causes the energy denominators 
to vary widely between channels, so the mass constraints assumed in the theorem are 
strongly violated.

\section{Numerical Results: Application to Charmonium}

\subsection{Mass Shifts}

To illustrate this formalism we will evaluate the effect
of open-charm meson loops on the masses and compositions 
of 1S, 1P and 2S charmonium states.
We use the well established ${}^3$P$_0$ model 
\cite{Ackleh:1996yt,Micu:1968mk,LeYaouanc:1972ae,LeYaouanc:1977ux,LeYaouanc:1977gm}
as the valence-continuum coupling Hamiltonian,
and neglect two-meson interactions. Our general approach is very 
similar to an earlier study by Heikkila, Tornqvist and Ono 
\cite{Heikkila:1983wd}, although we find somewhat larger loop effects 
than reported by this reference.

The ${}^3\P_0$ model treats strong decays as due to a bilinear
quark-antiquark pair production interaction Hamiltonian,
${\H}_{\I} = \gamma \sum_q 2 m_q \bar \psi_q \psi_q$, 
which is normally evaluated using nonrelativistic quark model matrix elements. 

A diagrammatic technique for determining
the valence-continuum coupling matrix element
$h_{fi}$ between a meson A and a two-meson state BC 
in the ${}^3\P_0$ model is given in Ref.\cite{Ackleh:1996yt}. 
We use this approach to determine the $\{ h_{fi}\} $ 
A-BC valence-continuum matrix elements. Gaussian momentum-space 
quark model meson wavefunctions were used, with unequal light, 
strange and charm quark masses. For simplicity a
common width parameter $\beta$ was assumed for all charmonium and open-charm 
meson wavefunctions; tests of the overlaps of more realistic Coulomb 
plus linear plus smeared hyperfine wavefunctions with Gaussians shows that this is 
a reasonable ``zeroth-order" approximation.

\subsubsection{$J/\psi$ mass shifts}

As a first numerical example we consider the mass shift 
of an initial valence $J/\psi$ $c\bar c$ state 
mixing with the DD continuum. 
The $h_{fi}$ matrix element for the transition
$J/\psi \to {\D}({\vec p}\, )\, \bar{\D}( -{\vec p}\, )$
for a rest $J/\psi$ in polarization state $m$ is given by
\be
h_{fi} = 
\frac{2^3 }{3^3 }\,
\frac{1+3r_n}{1+r_n}\,
\frac{\gamma}{\pi^{1/4} \beta^{1/2}}\;
\; \rho \, e^{- \rho^2 / 3(1+r_n)^2} 
{\rm Y}_{1m}(\Omega_p) 
\label{hfi_1}
\ee  
where $r_n = m_n/m_c$ is the light $(n=u,d)$ to charm quark mass ratio, 
$\rho =p/\beta$, $\beta$ is the simple harmonic oscillator (SHO) meson wavefunction width parameter 
(taken to be the same for all mesons in this work), 
and $\gamma$ is the dimensionless ${}^3$P$_0$ pair production amplitude. 
On substituting this $h_{fi}$ into the mass shift formula
Eq.(\ref{Mshift_pint}), and including a flavor factor of two for neutral and 
charged DD loops, we find 
\bd
\Delta \M_{J/\psi}^{({\D\D})} = \hskip 6cm
\ed
\be
-
\frac{2^7}{3^6}\,
\bigg(\frac{1+3r_n}{1+r_n}\bigg)^2
\frac{\gamma^2 \beta}{\pi^{1/2}}  
\int_0^{\infty} \!\!\!\!
\frac{\rho^4 e^{- 2 \rho^2 / 3(1+r_n)^2} d\rho }
{\big( 2 (\rho^2 + \mu_{\D}^2)^{1/2} - \mu_{J/\psi}\big) }
\hskip 1cm
\ee
where 
$\mu \equiv {\M}/\beta$ for each meson.

Numerical evaluation of this integral using
${\M}_{J/\psi} = 3.097$~GeV,
${\M}_{\D} = 1.867$~GeV,
$\beta = 0.5$~GeV,
$\gamma = 0.35$ 
(motivated by total widths; see Fig.2 of Ref.\cite{Barnes:2005pb})
and
$r_n = m_n/m_c = 0.33/1.5$ gives the result
\be
\Delta {\M}_{J/\psi}^{(\D\D)} = -23.1\ {\rm MeV}. 
\ee
Using the experimental $J/\psi$ mass as the input bare mass   
in this manner is of course only appropriate as an estimate 
of the size of these effects. Since this is in effect a renormalization
problem, the sum of the (unobservable) bare mass and mass shift 
should be identified with the experimental $J/\psi$ mass.

Although this DD contribution is a relatively small effect, 
incorporation of higher (1S)(1S) channels shows that the 
summed loop mass shifts are quite large. The formulas for the 
DD, DD$^*$ and D$^*$D$^*$ loop integrals in the $J/\psi$ system
are identical, but the relative spin-flavor factors of
1:4:7 give a combined mass shift that is an order of magnitude 
larger than for the DD channel alone.
(These 1:4:7 spin-flavor factors were
reported earlier by Heikkila {\it et al.} \cite{Heikkila:1983wd}
for loop contributions to mass shifts,  
and by De~Rujula {\it et al.} \cite{DeRujula:1976zg} and 
Close \cite{Close:1976ap}
for charm production cross sections.) 
On including all six 
D, D$^*$, D$_s$ and D$_s^*$ pair channels
(with 
${\M}_{\D}     = 1.867$~GeV,
${\M}_{\D^*}   = 2.008$~GeV,
${\M}_{\D_s}   = 1.968$~GeV,
${\M}_{\D_s^*} = 2.112$~GeV
and $r_s    = 0.55/1.5$), 
we find
\be
\sum_{n=1}^6\Delta {\M}_{J/\psi}^{(n)} 
= 
-457.5\ {\rm MeV},
\ee

This very large mass shift appears to invalidate the quenched quark model. In the next section
we will see that this scale of mass shift is actually common to all the low-lying charmonium states, 
and can therefore be approximately subsumed in a change of parameters 
(such as the charm quark mass $m_c$ or an overall constant $V_0$ in the $c\bar c$ potential).  

\subsubsection{Mass shifts of other charmonium states}

One can understand how such large mass shifts may have been accommodated 
in pure $c\bar c$ quark models
by evaluating the mass shifts of the remaining low-lying 
charmonium states below DD threshold.   
We again set the bare masses equal to the experimental values
to generate this estimate; the values used are 
${\M}_{\psi'}   = 3.686$~GeV,
${\M}_{\eta_c'} = 3.637$~GeV,
${\M}_{\chi_2}  = 3.556$~GeV,
${\M}_{\chi_1}  = 3.511$~GeV,
${\M}_{\chi_0}  = 3.415$~GeV,
${\M}_{h_c}     = 3.526$~GeV,
${\M}_{\eta_c}  = 2.979$~GeV,
and the other model parameters are as before.

\begin{table*}[h]
\caption{Mass shifts (in MeV) and $c\bar c$ probabilities 
for low-lying charmonium states 
due to couplings to two-meson continua. This 
one-loop estimate sets the unperturbed bare masses to the 
experimental values,
and assumes ${}^3\P_0$ model and SHO wavefunction parameters 
$\gamma = 0.35$ and $\beta = 0.5$~GeV and quark mass ratios
$r_n = m_n/m_c = 0.33/1.5$ and $r_s = m_s/m_c = 0.55/1.5$.}
\vskip 0.5cm
\begin{tabular}{lr|ccccccc|c} 
\hline
\multicolumn{2}{c|}{Bare $c\bar c$ State}
& 
\multicolumn{7}{c|}{Mass Shifts by Channel, $\Delta \M_i$  (MeV)}
&
\\
Multiplet
&
State\phantom{,,}
&\quad DD &\quad  DD$^*$ &\quad D$^*$D$^*$ 
&\quad  D$_s$D$_s$ &\quad  D$_s$D$_s^*$ &\quad D$_s^*$D$_s^*$ 
& \quad Total \quad & $\;$ P$_{c\bar c}$ \quad \\ 
\hline
1S  
&  $J/\psi(1^3{\S}_1) $ 
&  $-23$  & $-83$ & $-132$ 
&  $-21$  & $-76$ & $-123$ 
&  $-457$ 
&  0.69  \\
&  $\eta_c(1^1{\S}_0) $ 
&  $\phantom{-}0$  & $-114$ & $-105$ 
&  $\phantom{-}0$  & $-106$ & $-98$ 
&  $-423$ 
&  0.73  \\
\hline
2S 
&  $\psi'(2^3{\S}_1) $ 
&  $-27$  & $-84$ & $-126$ 
&  $-19$  & $-70$ & $-113$ 
&  $-440$ 
&  0.51  \\
&  $\eta_c'(2^1{\S}_0) $ 
&  $\phantom{-}0$  & $-118$ & $-103$ 
&  $\phantom{-}0$  & $-102$ & $-94$ 
&  $-416$ 
&  0.61  \\
\hline
1P &  $\chi_2(1^3{\P}_2) $ 
&  $-40$  & $-105$ & $-144$ 
&  $-33$  & $-88$ & $-111$ 
&  $-521$ 
&  0.49  \\
&  $\chi_1(1^3{\P}_1) $ 
&  $\phantom{-}0$  & $-127$ & $-148$ 
&  $\phantom{-}0$  & $-90$ & $-130$ 
&  $-496$ 
&  0.52  \\
&  $\chi_0(1^3{\P}_0) $ 
&  $-57$  & $\phantom{-}0$ & $-196$ 
&  $-34$  & $\phantom{-}0$ & $-172$ 
&  $-459$ 
&  0.58  \\
&  $h_c(1^1{\P}_1)    $ 
&  $\phantom{-}0$  & $-149$ & $-130$ 
&  $\phantom{-}0$  & $-118$ & $-107$ 
&  $-504$ 
&  0.52  \\
\hline
\hline
\end{tabular}
\label{Table_mass_shifts}
\end{table*}

The resulting mass shifts are given in Table~\ref{Table_mass_shifts}, 
and evidently are 
all quite large. Note however that they are rather similar, so
there is a much smaller scatter about the mean shift; the mean and 
variance are respectively $-471$~MeV and $49$~MeV. 
The scatter of mass shifts within a multiplet is even smaller;
the variance within the 1P multiplet for example is just $24$~MeV. 
(The similarity of mass shifts within a multiplet was discussed in the
previous section, and is a consequence of the general nature
of the valence-continuum coupling model.)

The large overall shift could be parameterized in a 
pure $c\bar c$ ``quenched" potential model
through a shift in $m_c$ or through the addition of a large negative 
constant $V_0$ to the $c\bar c$ potential. 
One expects that the goodness
of fit to the $c\bar c$ spectrum 
is rather insensitive to these modifications.  

The $J/\psi-\eta_c$ and $\psi'-\eta_c'$ loop-induced mass splitting has been discussed
previously by Eichten {\it et al.}\cite{Eichten:2004uh}. These authors sum over the same
set of intermediate states employed here, but use the Cornell decay model for the strong decay
interaction. They find a small loop-induced $J/\psi-\eta_c$ mass splitting of -3.7 MeV and a
$\psi'-\eta_c'$ splitting of -20.9 MeV, bringing their model into good agreement with the
experimental $\psi'-\eta_c$ mass difference. Table \ref{Table_mass_shifts} shows that we find
a numerically similar $\psi'-\eta_c'$ splitting of -24 MeV; however, the ground state mass difference
due to coupling to the continuum is -34 MeV, indicating that loop effects induce a larger 
$\psi'-\eta_c'$ mass difference of approximately +10 MeV. This is consistent with the bare model 
employed here, which 
finds a substantially smaller bare $\psi'-\eta_c'$ mass difference\cite{Barnes:2005pb} than that 
of Ref. \cite{Eichten:2004uh}.

\begin{table*}[h]
\caption{Relative one-loop mass shifts of 1P charmonium states
in the equal mass limit.}
\vskip 0.5cm
\begin{tabular}{c|ccc|ccc} 
\hline
Bare $c\bar c$ State 
& 
\multicolumn{6}{c}{Relative Mass Shifts,  
$\Delta {\M}_i({\rm L}_{\B\C})/\Delta {\M}_{tot}({\rm L}_{\B\C})$}
\\
\hline
& 
\multicolumn{3}{c|}{L$_{\B\C}$ = 0 }
&
\multicolumn{3}{c}{L$_{\B\C}$ = 2 }
\\
&\quad DD &\quad  DD$^*$ &\quad D$^*$D$^*$ 
&\quad DD &\quad  DD$^*$ &\quad D$^*$D$^*$ 
\\ 
\hline
$1^3{\rm P}_2 $ 
&  $0$  & $0$ & $1$ 
&  $3/20$  & $9/20$ & $2/5$ 
\\
$1^3{\rm P}_1 $ 
&  $0$  & $1$ & $0$ 
&  $0$  & $1/4$ & $3/4$ 
\\
$1^3{\rm P}_0 $ 
&  $3/4$  & $0$ & $1/4$ 
&  $0$  & $0$ & $1$ 
\\
$1^1{\rm P}_1    $ 
&  $0$  & $1/2$ & $1/2$ 
&  $0$  & $1/2$ & $1/2$ 
\\
\hline
\hline
\end{tabular}
\label{Table_rel_mass_shifts} 
\end{table*}

\begin{table*}[h]
\caption{Valence configuration mixing amplitudes $a_{fi}$ 
due to loops 
(DD; DD$^*$; D$^*$D$^*$ $^1$P$_1$; D$^*$D$^*$ $^5$P$_1$)
in the 
$
\{
|{}^3{\S}_1\rangle, |2{}^3{\S}_1\rangle, |{}^3{\D}_1\rangle \}
$
system.
The total $a_{fi}$ is the sum of the individual loop contributions, 
as indicated. The labels $|\I\rangle$ {\it etc.} refer to the physical 
(unnormalized) states one finds due to loop-induced mixing between 
$|c\bar c \rangle$ valence states.
(These are perturbative, one-loop results, with 
parameters as in Table~\ref{Table_mass_shifts}.)
Note the approximate cancellations in $\Delta\L \neq 0$ mixing, and the
non-symmetric mixing amplitudes.}
\vskip 0.5cm
\begin{tabular}{l|c|c|c}
\hline
&  $|\I\rangle$ 
&  $|\I\I\rangle$ 
&  $|\I\I\I\rangle$
\\
\hline
$|{}^3{\S}_1\rangle$
&  $ [1] $   
&  $ -.013 - .011 + .000 + .006 = -.018 $
&  $ -.089 - .017i + .086 - .010 - .020 = -.033- .017i $
\\
$|2{}^3{\S}_1\rangle$
&  $ -.003 -.014 -.001 -.026 = -.045 $ 
&  $ [1] $   
&  $ -.572 - .138i + .573 - .072 - .143 = -.214 - .138i $
\\
$|{}^3{\D}_1\rangle$
&  $ +.015 - .026 + .004 + .008 = +.001 $ 
&  $ +.340 - .469 + .063 + .126 = +.060 $ 
&  $ [1] $   
\\
\hline
\hline
\end{tabular}
\label{Table_afi} 
\end{table*}

\subsection{Continuum Components}

Although the large negative mass shifts may be ``hidden" by the choice 
of $m_c$ or $V_0$ in potential models and $m_c$ or $a(\beta)$ in 
quenched LGT, it should nonetheless be possible to identify other observable
effects of the two-meson continuum components, since according to Table~1 
their occupation probabilities are comparable to the valence $c\bar c$ 
components. To illustrate this we will evaluate some of these continuum 
component wavefunctions explicitly, and consider their effect on some 
experimentally observed properties of charmonium states. 

\begin{figure}[ht]
\vskip 0.7cm
\includegraphics[width=6cm,angle=270]{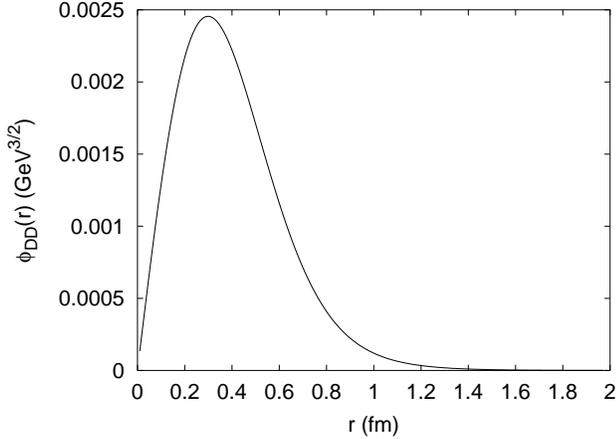}
\caption{DD Continuum Component Wavefunction of the $J/\psi$.}
\label{fig:DD}
\end{figure}

Recall from Eq.\ref{BCcontinuum_wfn} that the continuum 
component wavefunction in momentum space, $\phi_{\B\C}(\vec p \,)$, 
is given by
\be
\phi_{\B\C}(\vec p\, ) =
- \frac{h_{fi}}{({\E}_{\B\C}(p) - {\M}_{\A})} \ .
\ee 
Again specializing to the DD component of the $J/\psi$ as our example,
this momentum space wavefunction is 
\be
\phi_{\D\D}(\vec p \,) = 
\phi_{\D\D}(p) {\rm Y}_{1m}(\Omega_p)
\ee
where
\bd
\phi_{\D\D}(p) =  \hskip 6cm
\ed
\be
- \frac{8}{27}\,
\bigg(\frac{1+3r_n}{1+r_n}\bigg)\,
\frac{\gamma}{\pi^{1/4} \beta^{\, 3/2}}  
\frac{p \, e^{-  p^2 / 3(1+r_n)^2\beta^2}}
{\big( 2 ({\M}_{\D}^2 + p^2)^{1/2} - {\M}_{J/\psi} \big) } \ . 
\label{DD_wfn}
\ee  
Note that this component formally diverges as
${\M}_{J/\psi} \to 2 {\M}_{\D}$, due to a vanishing energy denominator;
this shows that as expected the largest continuum components 
arise in valence states that are closest to the continuum. 
The spatial wavefunction corresponding to
$\phi_{\D\D}(p)$ is shown in Fig. \ref{fig:DD}.

\subsection{$J/\psi$ continuum probabilities}

The probability of finding the physical $J/\psi$ 
in the DD continuum (from Eq.(\ref{BC_prob})) is
\bd
{\rm P}^{(\D\D)}_{J/\psi} = \hskip 6cm
\ed
\be
\frac{2^7}{3^6}\,
\bigg(\frac{1+3r_n}{1+r_n}\bigg)^2
\frac{\gamma^2}{\pi^{1/2}} 
\int_0^{\infty}\!\!\!
\frac{\rho^4 e^{- 2 \rho^2/ 3(1+r_n)^2} d\rho }
{\big( 2 (\rho^2 + \mu_{\D}^2)^{1/2} - \mu_{J/\psi}\big)^2} \ .
\ee
\bd
= 0.021 \ .
\ed
Although this appears to be a reassuring small correction to the valence
quark model description of the $J/\psi$ as a pure $c\bar c$ state, when we 
calculate the probability that the physical state is in any of
the $i=1\dots 6$ meson continuum states 
DD, D$^*$D, D$^*$D$^*$, 
D$_s$D$_s$, D$_s$D$_s$$^*$, D$_s$$^*$D$_s$$^*$,  
we again find that the summed contribution is quite large. Expressed as 
the probability that the physical $J/\psi$ is in the valence $c\bar c$ state,
we find
\be
{\P}^{(c\bar c )}_{J/\psi}
=
1 - \sum_{i=1}^6  {\P}^{(i)}_{J/\psi} 
= 0.685 \ .
\ee
Just as was the case for the mass shifts, we find that the 
continuum components of charmonium states are very large. This represents an interesting
challenge in the interpretation of the constituent quark model and quenched QCD, which both
neglect meson loops. The main issue is whether such large loop effects can be absorbed into
parameter redefinitions when computing observables.

\subsection{Spectroscopic Mixing}

As noted previously, discrete charmonium levels below the continuum
mix at second order in the valence-continuum 
Hamiltonian ${\H}_{\I}$ through hadron loops, provided that both 
the initial and final valence states $|i\rangle$ and $|f\rangle$ 
have nonzero matrix elements to at least one continuum intermediate 
state $|{\B\C}\rangle$. 

Here we shall illustrate this effect by calculating 
the amount of mixing between low-lying $1^{--}$ states. 
First we consider the 
$|{}^3{\S}_1 \rangle $ 
and  
$|2^3{\S}_1 \rangle $  
$c\bar c$ basis states, which at leading order are identified with the 
$J/\psi$ and $\psi(3686)$ respectively. 
We will give 
explicit formulas for mixing through DD intermediate states, 
and simply quote numerical results for mixing through 
higher two-meson continua.

The $h_{fi}$ matrix elements required to evaluate these mixing 
amplitudes are

\bd
h_{{\B\C},i}({}^3{\S}_1 \to {\D\D})  = \hskip 5cm
\ed
\be
\frac{2^3 }{3^3 }\,
\frac{1+3r_n}{1+r_n}\,
\frac{\gamma}{\pi^{1/4} \beta^{1/2}}\;
\; \rho \, e^{-  \rho^2 / 3(1+r_n)^2} 
{\rm Y}_{1m}(\Omega_p) \ ,
\label{hfi_1b}
\ee  

\bd
h_{f,{\B\C}}({\D\D} \to 2^3{\S}_1) =  \hskip 5cm
\ed
\bd
\frac{2^{5/2} }{3^{5/2} }\,
\bigg[ 
1 
+ 
\frac{2}{9}\frac{1}{(1+r_n)} 
- 
\frac{8}{27}\frac{ (1+3r_n)}{(1+r_n)^3}\, \rho^2 
\bigg]
\ed
\be
\cdot \;
\frac{\gamma}{\pi^{1/4} \beta^{1/2}}\;
\; \rho \, e^{- \rho^2 / 3(1+r_n)^2} 
{\rm Y}^*_{1m}(\Omega_p)  \ ,
\label{hfi_2}
\ee  
and 
\bd
h_{f,{\B\C}}({\D\D} \to {}^3{\D}_1) =  \hskip 5cm
\ed
\bd
\frac{2^{11/2} 5^{1/2} }{3^{9/2} }\,
\bigg[ 
\frac{r}{(1+r)} 
- 
\frac{2}{15}\frac{(1+3r_n)}{(1+r_n)^3}\, \rho^2 
\bigg]
\ed
\be
\cdot \;
\frac{\gamma}{\pi^{1/4} \beta^{1/2}}\;
\; \rho \, e^{- \rho^2 / 3(1+r_n)^2} 
{\rm Y}^*_{1m}(\Omega_p)  \ .
\label{hfi_3}
\ee  

Substitution of these expressions in Eq.(\ref{a_fi}) and 
evaluation of the overlap integral gives the $1^3{\S}_1$-$2^3{\S}_1$ 
mixing amplitudes $a_{fi}$. Our numerical results, using the same 
parameters and masses as previously, are given in Table~\ref{Table_afi}.

\section{Summary and Conclusions}

In this paper we presented a formalism for ``unquenching the
quark model" through the incorporation of the effects 
of hadron loops on valence quark model states. We gave expressions 
for the mass shift, continuum components of the hadron state vector,
and mixing amplitudes between discrete valence states that follow
from hadron loop effects for a given valence-continuum coupling Hamiltonian. 

As a numerical example we applied this formalism 
to the experimentally well-established light charmonium system, using the
$^3$P$_0$ decay model for the valence-continuum coupling.
We evaluated the mass shifts and composition of the physical
charmonium states for all 1S, 1P and 2S states using perturbation theory
in the valence-continuum coupling; these mass shifts
and two-meson components were found to be quite large. 
Since the mass shifts of the different charmonium levels are 
numerically rather similar, we speculate that they have been hidden 
in the choice of $m_c$ or a constant potential shift $V_0$
in $c\bar c$ valence potential 
models. It  remains to be seen whether the two-meson continuum components 
can be 
``parametrized away" -- it is possible that they lead to important 
mixing effects between discrete charmonium basis states that 
may be experimentally observable.

The mixing effects we find using the $^3$P$_0$ 
decay model as the valence-continuum coupling prove to be quite
large for higher-mass intermediate continuum states.  Although it is possible
that these effects can be largely renormalized away, an
accurate description of loop effects will probably 
require the development of a more realistic valence-continuum 
coupling Hamiltonian than the $^3$P$_0$ model.

\appendix*

\section{Loop Theorems}

Numerical experiments suggest that although individual loop contributions 
to physical observables are large, in practice there are often important 
cancellations or constraints when loop sums over sets of mesons are carried out.
This is evident for example in the mass shifts in Table~\ref{Table_mass_shifts};
the individual loop mass shift for a given state varies widely between states,
but the total mass shifts when summed over loops are rather similar. One can see 
that these relations are exact in certain limits. As an example, 
Table~\ref{Table_rel_mass_shifts} shows the relative mass shifts of the four P-wave 
charmonium states in the limit in which they have identical initial masses, and the 
D and D$^*$ within the loops also have identical masses; although the individual 
channel mass shifts differ, we find the same total mass shift for each P-wave state
on summing over the channels DD, DD$^*$ and D$^*$D$^*$. 

A similar result is evident in the loop-induced configuration mixing discussed in the text; 
the configuration mixing amplitude $a_{fi}$ between initial $i$ and final $f$ meson basis 
states in the usual $\N,\J,\L,\S$ basis is found to be zero if $\L_i\neq \L_f$, provided 
that the mesons in the loops have identical masses and we again sum over a complete 
set of loop meson spin states $\S_{\B}$ and $\S_{\C}$. As an example, in this limit this gives 
a zero mixing amplitude due to loops between any charmonium $^3\S_1$ and $^3\D_1$ basis states.

In this appendix we give a proof of this mass shift identity and the zero-mixing result
for loop sums; these results hold whenever one sums over loops containing a complete set of
spin (S) meson states (in a given N,J,L,S multiplet). The proof applies to the 
$^3$P$_0$ coupling model in particular, but also holds for a more general class of 
valence-continuum couplings, specifically to spin-one, factorised, spectator decay models, 
as discussed by Burns, Close, and Thomas \cite{Burns:2007hk}. 
In this type of model the valence-continuum coupling proceeds
through spin-one $q\bar q$ pair production, the initial quarks do not
couple to the decay vertex, and the spatial dependence
of the decay vertex multiplies the created $q\bar q$ spin operator:
${\cal O} = {\bf \sigma} {\bf \psi}$, where $\psi$ represents
the spatial portion of the decay vertex.
The proof therefore also applies
to the Cornell decay model\cite{Eichten:1975bk} and a decay model based on the nonrelativistic 
reduction of the interaction $\int \bar \psi \psi(\vec x) V(\vec x - \vec y) \bar \psi \psi(\vec y)$,
but does not apply to 
pair production from 
one gluon exchange (discussed in Ref.\cite{Ackleh:1996yt}).

Given a valence-continuum coupling of this general form, which includes the 
$^3$P$_0$ used in this paper for numerical examples, one may show that 
the general $\langle \B\C | H_I | \A \rangle$ matrix element is of the form

\begin{eqnarray}
&& \langle J_{\A} [L j_{\B\C}]; j_{\B\C} [j_{\B} j_{\C}]; j_{\B} [s_{\B} \ell_{\B}] \, 
j_{\C} [s_{\C} \ell_{\C}] | 
{\bm \sigma} {\bm \psi} 
| J_{\A} [s_{\A} \ell_{\A}]\rangle = \nonumber \\
&& \qquad \sum_{s_{\B\C} \ell_{\B\C} L_f} (-)^{\eta}
\hat{1} \hat{L}_f \hat{s}_{\B\C} \hat{\ell}_{\B\C} \hat{j}_{\B} \hat{j}_{\C} 
\hat{j}_{\B\C} \hat{s}_{\A} \hat{s}_{\B} \hat{s}_{\C} \hat{s}_{\B\C} \cdot \nonumber \\
&& \qquad \langle L_f [\L \ell_{\B\C}]; \ell_{\B\C} [\ell_{\B} \ell_{\C}] || {\bm \psi} 
|| \ell_{\A} \rangle \cdot 
\nonumber \\
&& \qquad \ninej{s_{\B}}{\ell_{\B}}{j_{\B}}{s_{\C}}{\ell_{\C}}{j_{\C}}{s_{\B\C}}
{\ell_{\B\C}}{j_{\B\C}}
\ninej{1/2}{1/2}{s_{\B}}{1/2}{1/2}{s_{\C}}{s_A}{1}{s_{\B\C}} \cdot \nonumber \\
&& \qquad \sixj{s_{\B\C}}{\ell_{\B\C}}{j_{\B\C}}{\L}{j_{\A}}{L_f}
\sixj{s_{\B\C}}{s_{\A}}{1}{\ell_{\A}}{L_f}{j_{\A}} 
\label{9jEq}
\end{eqnarray}
where $\hat x = \sqrt{2x+1}$ and $\eta = \L + s_{\B\C} + \ell_{\B\C} + L_{f} + s_{\B}$.

If the expressions for the 
mass splitting (Eq.\ref{Mshift_pint})
or spectroscopic mixing are
summed over intermediate states BC with identical masses, the resulting 
common energy denominators may be taken outside the sum over channels, 
and one is left with the expressions

\begin{equation}
\delta m(i) = \int \frac{d^3p}{(2\pi)^3}
\frac{1}{(m_i - {\bar{\E}}_{\B\C}(p) + i \epsilon )}
\sum_{\B\C} |\langle j_i [s_i \ell_i] | {\bm \sigma} {\bm \psi} | \B\C\rangle |^2
\label{mEq}
\end{equation}
and
\begin{eqnarray}
a_{fi} &=& \frac{1}{(m_i-m_f)} \int \frac{d^3p}{(2\pi)^3}
\frac{1}{(m_i - {\bar{\E}}_{\B\C}(p) + i \epsilon )} \cdot \nonumber \\
&& \sum_{\B\C}
\langle j_f [s_f \ell_f] | {\bm \sigma} {\bm \psi} | \B\C\rangle
\langle \B\C | {\bm \sigma}{\bm \psi}|j_i [s_i \ell_i]\rangle
\label{aEq}
\end{eqnarray}
where 
${\bar{\E}}_{\B\C}$ 
is the common energy of all states in the same multiplet 
as BC.

The sum over intermediate states simplifies when one considers a subsum 
over spin multiplets:
\begin{equation}
\sum_{\B\C} \to \sum_{s_{\B} s_{\C} j_{\B} j_{\C}};
\end{equation}
the angular momenta $\ell_{\B}$, $\ell_{\C}$, $\L$ can remain fixed.

On substituting Eq.\ref{9jEq} into Eqs.\ref{mEq} and \ref{aEq}, 
and using the orthogonality relation for $9j$ and $6j$ symbols

\begin{eqnarray}
&& \sum_{j_{13},j_{24}} \hat{j}_{13}\hat{j}_{24} \ninej{j_1}{j_2}{j_{12}}{j_3}{j_4}{j_{34}}{j_{13}}{j_{24}}{J} 
\cdot \ninej{j_1}{j_2}{j_{12}'}{j_3}{j_4}{j_{34}'}{j_{13}}{j_{24}}{J} = \nonumber \\
&& \frac{\delta(j_{12}, j_{12}') \delta(j_{34}, j_{34}') } {\hat{j}_{12} \hat{j}_{34} }
\end{eqnarray}
and
\begin{equation}
\sum_{j_{12}} \hat{j}_{12}^2 \sixj{j_1}{j_2}{j_{12}}{j_3}{j_4}{J} \cdot \sixj{j_1}{j_2}{j_{12}}{j_3}{j_4}{J'} = \frac{\delta(J, J')}{\hat{J}},
\end{equation}
we obtain the following sum:
\begin{equation}
\frac{\delta_{s_i s_f} \delta_{\ell_i \ell_f}}{2 \ell_i+1} \sum_{\ell_{\B\C} L_f} |
\langle L_f [\L \ell_{\B\C}]; \ell_{\B\C} [\ell_{\B} \ell_{\C}] ||
{\bm \psi}
|| \ell_i\rangle |^2.
\end{equation}

Since this expression is independent of the initial and final meson spin,
we conclude that all mesons in a given (assumed degenerate) 
spin multiplet receive the same width and mass shift from 
the sum over all intermediate (loop) mesons in a given spin multiplet. 
Furthermore, the spectroscopic mixing
between mesons of different orbital angular momentum is 
zero when sums over spin multiplet intermediate states are carried out. 
(The external meson masses need not be identical to prove this result.) 
Finally, since these matrix elements drive nonperturbative mixing 
(see the discussion following Eq.\ref{eq:bubble_sum}), 
these conclusions also apply to nonperturbative mixing, 
in the absence of final state interactions.

Spectroscopic mixing between mesons with differing
radial quantum numbers (but identical otherwise) is not zero in general.
The size of this mixing is governed by the spatial dependence of 
the strong decay vertex.
Spectroscopic mixing has been studied previously by 
Geiger and Isgur\cite{Geiger:va}, who considered the closure approximation, 
in which {\it all} loop mesons are assumed to be degenerate,
not simply those in a spin multiplet. 
Geiger and Isgur used this approximation to explain the observed 
weakness of loop-driven OZI violation effects.
We remark that the closure approximation implies that spectroscopic mixing between states
with different radial quantum numbers is zero:
under this approximation Eq.\ref{aEq} simplifies to
\begin{eqnarray}
a_{fi} &=& \frac{1}{(m_i-m_f)} \frac{1}{(m_i - \bar{\E})}\;
\langle n_f j_f [s_f \ell_f] | {\cal O}^2 | n_i j_i [s_i \ell_i\rangle \nonumber \\
&=& \frac{\langle 0 | {\cal O}^2| 0 \rangle}{(m_i-m_f)(m_i - \bar{\E})}\;
\langle n_f j_f [s_f \ell_f] |n_i j_i [s_i \ell_i] \rangle \nonumber \\
\end{eqnarray}
where the last form follows from the spectator nature of the decay model.
Thus, if we impose the equality of all loop meson masses,
{\it all} spectroscopic mixing is zero in the $^3$P$_0$ model, and in a 
wide range of related decay models.

Finally, the previous discussion remains largely unchanged when considering mixing 
between initially degenerate states. In this case one must diagonalize the matrix of second 
order matrix elements in the degenerate subspace\cite{vV}, $\delta H_{ij} = (m_i-m_j) a_{ji}$. 
Under the conditions of the theorem, off-diagonal matrix element in $\delta H_{ij}$ 
are zero when the meson spins or angular momenta differ. Furthermore the
diagonal matrix elements are identical. Thus conclusions concerning  mass shifts, widths, 
and small or zero spectroscopic mixing remain unchanged.

\acknowledgments

We acknowledge useful communications
with E.~van~Beveren, T.~Burns, S.~Capstick, K.~T.~Chao,
F.~E.~Close, S.~Godfrey, T.~Papenbrock, C.~Quigg, J.-M. Richard, J.~Rosner 
and C.~Y.~Wong in the course of this work.
ESS acknowledges support from the 
Rudolph Peierls Centre for Theoretical Physics,
Oxford University, where some of this work was carried out.
This research was supported in part by the U.S. National Science
Foundation through grant NSF-PHY-0244786 at the University of Tennessee,
the U.S. Department of Energy under contracts
DE-AC05-00OR22725 at Oak Ridge National Laboratory and
DE-FG02-00ER41135 at the University of Pittsburgh,
and by PPARC grant PP/B500607 at Oxford.

\end{document}